\documentstyle[twocolumn,eqsecnum,aps,prb,epsf]{revtex}
\def\mB{$\mu_{\rm B}$\ }
\def\veck{\mbox{\boldmath $k$}}
\draft
\preprint{}
\begin{document}
\title{Effect of Crystal Structure to Resonant X-Ray Scattering on YTiO$_3$}

\author{Manabu Takahashi and Jun-ichi Igarashi}
\address{Faculty of Engineering, Gunma University, Kiryu, Gunma 376-8515, Japan}

\date{\today}

\maketitle
\widetext
\begin{abstract}
We investigate the mechanism of the resonant X-ray scattering on
the Ti $K$ edge region of YTiO$_3$ using the band structure calculation
combined with the local density approximation.  A large intensity
is obtained for the orbital
superlattice spots.  The calculated spectra consist of several
peaks as a function of photon energy in agreement with the recent
experiments.  Against a naive interpretation that directly relates
the intensity to the orbitally polarized $3d$ states, the obtained
large intensity arises from the distorted crystal structure, {\it
i.e.}, the tilt of the TiO$_6$ octahedra and the Jahn-Teller
distortion, which considerably modifies the $4p$ states in the
intermediate states of the dipolar process.  This casts doubt on
a prevailing assertion that the resonant x-ray scattering is a direct 
observation of the orbital order.
\end{abstract}
\narrowtext
\section{Introduction}
Recently the resonant x-ray scattering spectroscopy has attracted
much attention because this experimental method is considered as
one of the most powerful tools which can directly observe the
orbital order in the strongly correlated systems.  A lot of
experiments have already been carried out on the perovskite
transition-metal compounds,
\cite{Murakami98a,Murakami98b,murakami99,murakami99b,murakami99c,murakami00b,Nakao00}
V$_2$O$_3$\cite{vettier99} and DyB$_2$C$_2$,
\cite{murakami00} in which the scattering intensities were observed
on the orbital superlattice spots, which are forbidden for usual
x-ray scattering.  Such anomalous x-ray scattering will simply be
referred to as RXS.  Despite much effort so far, how and to what
extent the RXS spectra are reflecting the orbital order are still
controversial, especially on the perovskite compounds, since the
orbital order usually accompanies the Jahn--Teller distortion (JTD)\cite{Kanamori}
and the GdFeO$_3$ type distortion, which can also gives rise to
the RXS intensity.

The resonant scattering may be described by a second order process
that a photon is virtually absorbed by exciting a core electron to
unoccupied states and then emitted by recombining the excited
electron with the core hole. Therefore, for the $K$ edge of the
transition-metal compounds, the $4p$ states are involved in the
intermediate states of the dipolar process.  For the RXS in the
$K$ edge of LaMnO$_3$, the so called {\it Coulomb mechanism} was
proposed that the anisotropic part of the intra-atomic Coulomb
interaction between the orbitally polarized $3d$ states and the
$4p$ states is responsible to the RXS intensity by making the
degenerate $4p$ levels split.\cite{Ishihara}  
However, subsequent studies based
on the band structure calculations have revealed that the Coulomb
effect is much smaller ($<1/100$) than the JTD effect on the RXS
intensity.\cite{Elfimov,Benfatto,Taka1}
This is because the $4p$ states are so extended in
space that they are sensitive to neighboring O potentials.

\newpage
~~~
\vskip 150pt
The RXS spectra on YVO$_3$ have been observed for the pre $K$ edge
and the main $K$ edge regions of V.\cite{murakami00b} The spectra
consist of several peaks as a function of photon energy, and the
shapes are different between the $(100)$ and $(011)$ reflections.
Assuming that the scattering tensor is directly related to the
orbital polarization of the $3d$ states, the authors argued
that the azimuthal angle dependence of the spectra 
was consistent with the C-type orbital order\cite{Sawada98} in the low temperature phase.  
However, this assumption made there implies the
dominance of the Coulomb effect, which is questionable.  Recently
the resonant x-ray scattering experiment has also been carried out
on YTiO$_3$, and the RXS spectra have been observed with their
shapes similar to those of YVO$_3$.\cite{Nakao00} In this paper,
concentrating our attention on YTiO$_3$, we study the mechanism of
the RXS spectra by carrying out a band structure calculation on
the experimentally determined crystal structure.\cite{Daivid79}

\begin{figure}
\begin{center}
\leavevmode
\epsfxsize=7cm \epsfbox{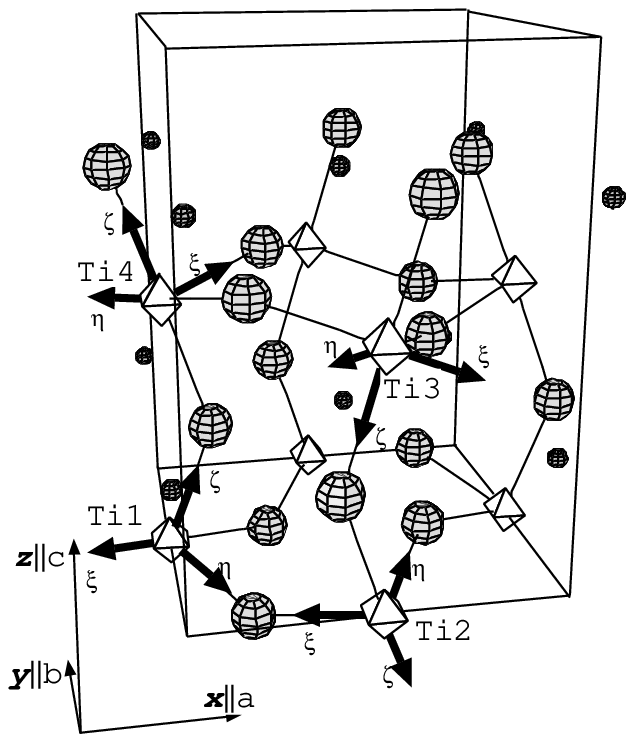}
\end{center}
\caption{Sketch of the crystal structure of YTiO$_3$.  The large
spheres represent the oxygen ions, the small solid spheres the
yttrium ions, and the octahedra the titanium ions. The arrows
assigned to each Ti ion define the local $\xi\eta\zeta$ axes pointing
to nearest neighbor oxygen sites.  }
\label{crystal} \end{figure}
The crystal structure of YTiO$_3$ is known to belong to the space
group $Pbnm$, as shown in Fig.~\ref{crystal}; four Ti ions are
inequivalent in the unit cell.  The Ti-O-Ti bond angles strongly
deviate from $180^\circ$ due to the tilt of the TiO$_6$ octahedra,\cite{com1}
and Ti-O bond lengths are different in TiO$_6$ octahedra due to
the JTD.  One of the $t_{2g}$ levels is occupied under the crystal
field in the $d^1$ configuration of Ti$^{3+}$ ions, and since the
$t_{2g}$ states are weakly hybridizing with O $2p$ states, the JTD
is smaller ($\sim 20\%$) than that on LaMnO$_3$\cite{Daivid79}.  Note that the
crystal structure of YVO$_3$ belongs to the same space group of
$Pbnm$ with similar sizes of the tilt of VO$_6$ octahedra and the
JTD\cite{Sawada98,Kawano94}. 

We calculate the RXS spectra using the band calculation based on
the local density approximation (LDA) together with the muffin-tin
approximation.  This approximate scheme leads to a metallic ground
state for YTiO$_3$ with a very small orbital polarization.  In spite
of this shortcoming for the $3d$ bands, we expect that the $4p$
bands are well described in the present calculation, since they
have energies $\sim 15$ eV higher than the $3d$ bands and thereby
the details of the $3d$ bands are irrelevant.  What is important,
which will become clear, to make the $4p$ states different at
different Ti sites is the O potential of neighboring sites.  The
$4p$ states are so extended in space that they are sensitively
modified by hybridizing and interacting with the electronic states
of neighboring ions.  Actually, for LaMnO$_3$, it is known that
the LDA$+U$ method, which predicts an insulating ground state with
large orbital polarization, gives the $4p$ bands and the RXS spectra
nearly the same as those given by the LDA method, although the latter
predicts a small band gap and a small orbital polarization.\cite{Benedetti00}

We obtain the RXS spectra with sufficient intensities for YTiO$_3$.
The obtained spectra as a function of photon energy
resemble closely to the results of the experiment for the $(100)$
and $(011)$ reflections of YVO$_3$,\cite{murakami00b}
and also to those of the recent experiment of YTiO$_3$.\cite{Nakao00}
Since the muffin-tin approximation averages the Coulomb interaction
between the $4p$ and the $3d$ states, the anisotropy of the Coulomb 
interaction is eliminated in the calculation of the RXS spectra.
Therefore the obtained RXS intensity arises from the distorted crystal 
structure.
We may roughly estimate that the Coulomb effect is less than $1/4$
of the JTD effect, since the Coulomb effect is evaluated as 1/100
smaller than the JTD effect in LaMnO$_3$.\cite{Benfatto,com2} Note
that the JTD effect cannot be distinguished from that of the tilt
of the TiO$_6$ octahedra in the present calculation.  Such sensitivity
of the $4p$ states to the electronic states at neighboring sites
has already been recognized in the analysis of the magnetic circular
dichroism of the $K$ edge absorption spectra on the ferromagnetic
states of transition metals\cite{Iga1} and the analysis of the
resonant x-ray magnetic scattering spectra on CoO\cite{Iga2} and NiO.\cite{Iga3}

This paper is organized as follows.  In the next section, we briefly
discuss the band structure calculation on the ground state.  In
section 3 we outline the calculation procedure of RXS spectra and
discuss the calculated results.  The final section
is devoted to concluding remarks.

\section{Calculation of the ground state and density of states}

\begin{figure}
\begin{center}
\leavevmode
\epsfxsize=7cm \epsfbox{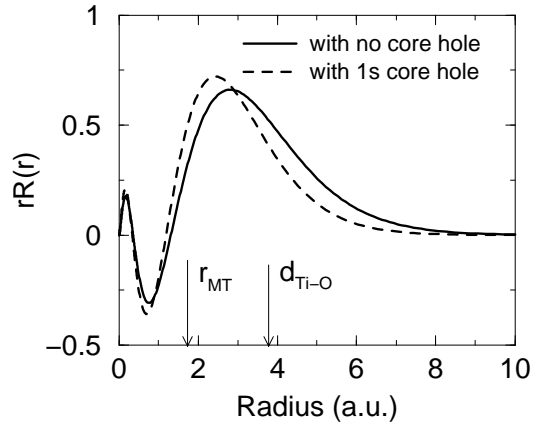}
\end{center}
\caption{The $4p$ wave functions of an atomic Ti$^{3+}$ ion, calculated
within the Hartree-Fock approximation.  Solid line and dashed line
represent the radial wave functions $rR(r)$ without and with the
core--hole potential.  The arrows indicate the muffin-tin radius
$r_{\rm MT}$ of Ti site and the typical distance $d_{\rm Ti-O}$
between Ti nucleus and O nucleus in YTiO$_3$.}
\label{wavefunc} \end{figure}
First we examine the meaning of the atomic $4p$ states.  Figure
\ref{wavefunc} shows the atomic $4p$ wave function for a Ti$^{3+}$
ion, calculated within the Hartree-Fock approximation.\cite{Cowan}
The $4p$ wave function extends considerably in space so that a
large part of the amplitude would lie on the neighboring oxygen
sites or the interstitial region in solids.  This situation is not
changed in the presence of the core-hole potential, as shown by
the broken line in the figure. Thus one may easily understand the
sensitiveness of the $4p$ states to electronic states at neighboring
sites.  The concept of the atomic $4p$ levels loses its clear
meaning in solids, and the $4p$ states are to be treated as energy
bands.  Nevertheless, the $p$ symmetric states (centered at a Ti site)
above the Fermi energy are referred to the $4p$ states, since
the atomic character is almost preserved inside the atomic sphere
in the band structure calculation.

In the band structure calculation for YTiO$_3$, we use the
Korringa--Kohn--Rostoker (KKR) method within the LDA scheme.  We
assume the crystal structure determined experimentally at room temperature
in Ref.~\cite{Daivid79}, which is shown in Fig.~\ref{crystal}.  
As mentioned before, the crystal structure
belongs to the space group $Pbnm$;\cite{Daivid79} the unit cell
involves four Y ions, four Ti ions and 12 O ions.  The four Ti
sites are distinctive in terms of the orientation even though the
orbital ordering does not occur.  The large tilt of TiO$_6$ octahedra
makes the Ti-O-Ti bond angles strongly deviates from $180^\circ$
by as much as $35^\circ$.  Each TiO$_6$ octahedron also undergoes
the JTD such that the direction $\xi$ corresponds to the short bond,
the $\eta$ to the long bond, and the $\zeta$ to the middle bond,
where three arrows $\xi\eta\zeta$ attached to each Ti ion are
directed to neighboring oxygen sites and are almost orthogonal to
each other.  There exists an important symmetry that the configuration
surrounding the Ti 1 site becomes coincident to that surrounding
the Ti 2 site under the rotation of all positions of atoms by $\pi$
around the line parallel to the $x$ axis passing through the Ti 1
site.  In the same way, the configuration surrounding the Ti 1 site
becomes coincident to those surrounding the Ti 3 and the Ti 4 sites
under the rotation of the atomic position by $\pi$ around the line
parallel to the $y$ and the $z$ axes, respectively, passing through
the Ti $1$ site.

As regards the magnetism, the system undergoes a ferromagnetic
transition below the Curie temperature $T_{\rm C}=29$K.\cite{Goral82}
Neglecting the spin-orbit coupling, we obtain a ferromagnetic ground state
with the magnetic moment $0.95$\mB per Ti$^{3+}$ site 
($0.68$\mB inside the muffin-tin sphere of Ti$^{3+}$ site), 
consistent with the previous calculations.\cite{Sawada98,asano95} 
This value is a little larger than the experimental value $0.84$\mB.\cite{Goral82}
The orbital polarization is underestimated to be very small.
The ground state is obtained as a metal, although it is experimentally
a semiconductor.  
The LDA is known to have a tendency of wrongly predicting a metallic 
ground state.
We think these shortcomings for the $3d$ states have only minor effects
on the $4p$ states, since they have high energy ($\sim 15$ eV above
the Fermi level).

\begin{figure}
\begin{center}
\leavevmode
\epsfxsize=7cm \epsfbox{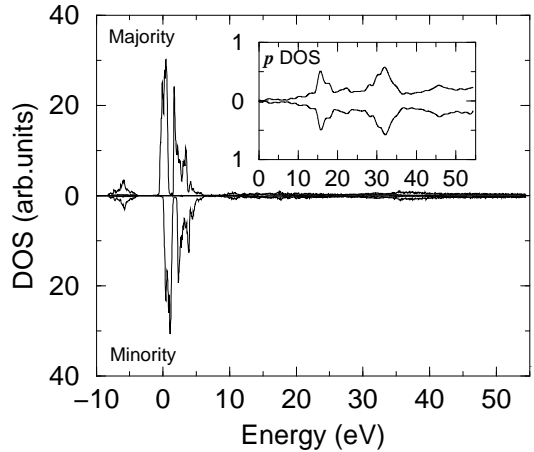}
\end{center}
\caption{The DOS projected onto the $s$, $p$ and $d$ symmetric
states inside the muffin-tin sphere of a Ti site. The origin of
energy is at the Fermi level.  The $d$ DOS is overwhelming
the $s$ and $p$ DOS's.  The inset shows the magnified DOS projected
onto the $p$ symmetric states.  }
\label{dosall} \end{figure}
Figure \ref{dosall} shows the density of states projected onto the
$s$, $p$, and $d$ symmetric states inside the muffin--tin sphere
of a Ti site.  Around $7$ eV below the Fermi level, the $3d$ states
form bonding states with the $2p$ states of the neighboring O
ions.  Around the Fermi level, the $3d$ states are highly concentrating
in a form of antibonding states with the O $2p$ states.  
In high energy region, a broad band appears, to which the $s$, $p$ and 
$d$ symmetric states equally contribute.  
The $p$ DOS in this energy region is much
smaller than the $d$ DOS near the Fermi level, 
as expected from the character of the
atomic $4p$ wave function that extends considerably in the interstitial
region. The $p$ DOS is enlarged in the inset. It has two peaks.
Note that the $p$ DOS is roughly proportional to the $K$ edge absorption 
spectrum, if the effect of the core hole potential is neglected
in the final states.  This final state interaction may enhance the peak 
of the low energy side in the $p$ DOS.  
In any event, the two-peak structure corresponds well to the absorption 
spectrum of YVO$_3$.\cite{murakami99b}

\section{Calculation of RXS spectra}
\subsection{Formulation}
In the resonant process around the Ti $K$ edge region, the $1s$
core electron is virtually excited to the $4p$ states. Therefore
the RXS spectra directly reflect the $4p$ states at the scattering
center.  We specifically calculate the RXS intensities for the
(100), (001) and (011) reflections.

We choose the $x$, $y$ and $z$ axes such that they are parallel to
the crystal a, b, and c axes, respectively.  With respect to these
axes, we define the $1s$--$4p$ dipole transition density matrix 
$\tau^{(k)}_{mn}(\varepsilon)$ at the Ti $k$ site in the unit cell as
\begin{eqnarray}
&&\tau^{(k)}_{mn}(\varepsilon) = \sum_{b,\veck}\int r^2 dr\, r'^2 dr' 
     [R_{1s}^*(r ) r  {\cal P}_m\phi_{b,\veck}(r )]^* \nonumber \\
&& ~~~~~~~~~~~~~~~~~~~  [R_{1s}^*(r') r' {\cal P}_n\phi_{b,\veck}(r')]
     \delta(\varepsilon-\varepsilon_{b,\veck}),
\end{eqnarray}
where $\phi_{b,\veck}$ represents the wave function with the band index
$b$, wave-vector {\boldmath $k$} and energy  $\varepsilon_{b,\veck}$.
The ${\cal P}_m$ is a projection operator which projects the wave
function $\phi_{b,\veck}$ onto the $m$ ($m\in x,y,z$) component of
the $p$ symmetric part around the Ti $k$ site,
and $R_{1s}$ represents the Ti $1s$ wave function.  
This matrix is obtained from the KKR Green's function
$g^{(k)}_{mn}(r,r';\varepsilon)$\cite{zeller} by the equation
\begin{equation}
\tau^{(k)}_{mn}(\varepsilon) = -\frac 1\pi {\rm Im} \int r^2 dr\, r'^2 dr'\, g^{(k)}_{mn}(r,r';\varepsilon) rr' R_{1s}(r) R_{1s}(r').
\end{equation}
Numerically this $1s$--$4p$ transition
density matrix is roughly proportional to the $4p$ density matrix
$\rho^{(k)}_{mn}(\varepsilon)= -\frac 1\pi {\rm Im} \int_0^{r_{MT}}
r^2 dr\, g^{(k)}_{mn}(r,r;\varepsilon)$ with $r_{\rm MT}$ being 
the muffin-tin radius. Assuming that photons
are linearly polarized, we have the scattering amplitude at the Ti $k$ site
as
\begin{equation}
f^{(k)}(\omega) = \frac A4 \int d\varepsilon
                  \sum_{m,n\in x,y,z}
                  E^{\mbox{\scriptsize out}}_m 
                      \frac{\tau^{(k)}_{mn}(\varepsilon)}
                           {\omega-\varepsilon+\varepsilon_{1s}+{\rm i}{\mit\Gamma}}
                  E^{\mbox{\scriptsize in}}_n,
\end{equation}
where $A$ is a numerical constant, $E^{\mbox{\scriptsize in}}_m$ and
$E^{\mbox{\scriptsize out}}_n$ are the $m$ and $n$ components of
polarization of the incident and the emitted photons, respectively.
In the denominator, $\omega$ is the photon energy, and $\varepsilon_{1s}$ 
is the energy of the $1s$ state. The $\mit\Gamma$ describes the broadening 
due to the core--hole lifetime, which are assumed ${\mit\Gamma}=1$ eV.
The effect of the core-hole potential is neglected in the intermediate states. 
We think this effect on the RXS intensity is small, since 
the core hole potential is spherically symmetric, and 
it affects merely the diagonal elements 
of the density matrix which are irrelevant to the RXS intensity.
The integration in terms of $\varepsilon$ is
performed from the Fermi energy to $60$ eV above it.

It is obvious that the transition density matrix $\tau^{(k)}_{mn}$ becomes
symmetric with neglecting the spin--orbit interaction.\cite{Iga2,Iga3}
Using $a$, $b$, $c$, $\alpha$, $\beta$ and $\gamma$, functions of 
$\varepsilon$, the matrix at the Ti 1 site can be represented as
\begin{equation}
\tau^{(1)} = \left( \begin{array}{rrr} a         & \alpha   & \gamma \\
                                       \alpha    &  b       & \beta  \\
                                       \gamma    & \beta    & c      
                    \end{array} \right).
\end{equation}
Due to the crystal symmetry mentioned before, 
the transition density matrices
$\tau^{(2)}$, $\tau^{(3)}$ and $\tau^{(4)}$ at the Ti $2$, Ti $3$,
and Ti $4$ sites are represented as 
\begin{eqnarray}
\tau^{(2)}&=&\left( \begin{array}{rrr} a         &-\alpha   &-\gamma \\
                                      -\alpha    &  b       & \beta  \\
                                      -\gamma    & \beta    & c      
                    \end{array} \right), \nonumber \\
\mbox{\ \ \ }
\tau^{(3)}&=&\left( \begin{array}{rrr} a         &-\alpha   & \gamma \\
                                      -\alpha    &  b       &-\beta  \\
                                       \gamma    &-\beta    & c      
                    \end{array} \right), \nonumber \\
\mbox{\ \ \ }
\tau^{(4)}&=&\left( \begin{array}{rrr} a         & \alpha   &-\gamma \\
                                       \alpha    &  b       &-\beta  \\
                                      -\gamma    &-\beta    & c      
                    \end{array} \right).
\end{eqnarray}
Using these representations, the scattering amplitudes are expressed
in simple forms. For the (100) reflection, we have
\begin{eqnarray}
&&f_{(100)}(\omega) = f^{(1)}(\omega)- f^{(2)}(\omega)-f^{(3)}(\omega)+f^{(4)}(\omega) \nonumber \\
                  &&=A (E^{\mbox{\scriptsize out}}_x E^{\mbox{\scriptsize in}}_y + E^{\mbox{\scriptsize out}}_y E^{\mbox{\scriptsize in}}_x )
		       \int d\varepsilon\frac{\alpha(\varepsilon)}{\omega-\varepsilon+\varepsilon_{1s}+{\rm i}{\mit\Gamma}}.
\label{f100} \end{eqnarray}
In the similar way, we have the scattering amplitude for the (001) and
(011) reflections as
\begin{eqnarray}
&&f_{(001)}(\omega) = A  (E^{\mbox{\scriptsize out}}_y E^{\mbox{\scriptsize in}}_z + E^{\mbox{\scriptsize out}}_z E^{\mbox{\scriptsize in}}_y) \nonumber \\
&& \int d\varepsilon\frac{\beta (\varepsilon)}{\omega-\varepsilon+\varepsilon_{1s}+{\rm i}{\mit\Gamma}}, \label{f001} \\
&&f_{(011)}(\omega) = A  (E^{\mbox{\scriptsize out}}_x E^{\mbox{\scriptsize in}}_z + E^{\mbox{\scriptsize out}}_z E^{\mbox{\scriptsize in}}_x) \nonumber \\
&& \int d\varepsilon\frac{\gamma(\varepsilon)}{\omega-\varepsilon+\varepsilon_{1s}+{\rm i}{\mit\Gamma}}. \label{f011}
\end{eqnarray}
Only one of the off-diagonal elements remains in each expression
of the scattering amplitudes, with the diagonal elements $a$, $b$, and
$c$ completely vanishing.  Therefore, we can select out one of
them by choosing the reflection index. The photon energy dependence
is factored out from the polarization of photon in Eqs.~(\ref{f100}),
(\ref{f001}), and (\ref{f011}).  The RXS intensities $I_{(100)}$,
$I_{(001)}$, and $I_{(011)}$ are given by $|f_{(100)}|^2$,
$|f_{(001)}|^2$, and  $|f_{(011)}|^2$, respectively.

\subsection{Calculated results}

We calculate the RXS spectra on the crystal structure determined 
at room temperature in Ref.~\cite{Nakao00}, using the above formulas.
We show in the following the results calculated in the ferromagnetic 
ground state of YTiO$_3$, 
although the RXS experimental data are given above the Curie temperature 
$T_{\mbox{c}}$.\cite{Nakao00} 
We checked that the RXS spectra are insensitive to the magnetic states
by calculating the spectra also in a nonmagnetic state.
This is consistent with the $4p$ DOS above $15$ eV 
in Fig. \ref{dosall}, which shows no exchange splitting.

\begin{figure}
\begin{center}
\leavevmode
\epsfxsize=7cm \epsfbox{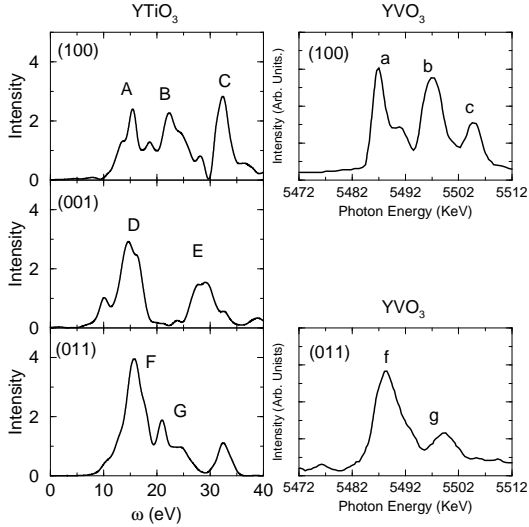}
\end{center}
\caption{RXS spectra as a function of photon energy.
The left panels are the calculated spectra for the (100), (001), and
(011) reflections on YTiO$_3$.
The origin of energy corresponds to the photon energy of exciting an
electron from the $1s$ state to the Fermi level.
Right panels are the experimental spectra in the low temperature
phase of YVO$_3$ (ref. 6).
}
\label{RXSall} \end{figure}
Figure \ref{RXSall} shows the calculated spectra as a function of photon 
energy for the (100), (001) and (011) reflections.
Note that the anisotropy of the Coulomb interaction between the $4p$ and
the $3d$ states is eliminated in the calculation by using the muffin-tin
approximation.
The experimental spectra in the low temperature phase of YVO$_3$ 
are also shown for the sake of comparison,\cite{murakami00b}
since their shapes are quite similar to the unpublished experimental data
for the RXS spectra of YTiO$_3$.\cite{Nakao00}
As already mentioned, the resemblance of the RXS spectra is quite 
understandable in the light of the mechanism of the distorted crystal structure,
since both the JTD and the tilt of the VO$_6$ octahedra in the low temperature
phase of YVO$_3$ are quite similar to those in YTiO$_3$.
In the (100) and (001) reflections,
the intensities vanish for the $\sigma\to\sigma'$ channel.
For the $\sigma\to\pi'$ channel, we have three peaks denoted by A, B, and C,
whose intensities are comparable to each other in the (100) reflection,
while we have two peaks denoted as D and E in the (001) reflection.
In the experimental data of YTiO$_3$, three peaks appear at 4.974 keV,
4.986 keV, and 5.0 keV, corresponding to peaks A, B, C, for
the (100) reflection, while two peaks appear at 4.974 keV and 4.984 keV
with the former intensity about twice as the latter, corresponding to
peaks D and E.\cite{Nakao00} In the (011) reflection, both the 
$\sigma\to\sigma'$ and the $\sigma\to\pi'$ channels contribute to 
the intensities;
we have a dominant peak denoted as F and a small peak denoted as G.
The energy of peak F is found slightly higher ($\sim 2$ eV) than that of
the peaks A and D. In the experimental data of YTiO$_3$, two peaks appear
at 4.976 keV (2 eV higher than the positions of peaks A and D)
and 4.986 keV with the former intensity dominating the latter.

\begin{figure}
\begin{center}
\leavevmode
\epsfxsize=6cm \epsfbox{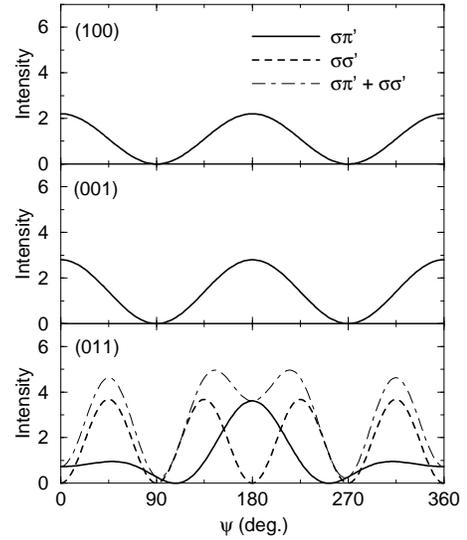}
\end{center}
\caption{Azimuthal angle dependence of the RXS intensities
for the (100), (001) and (011) reflections.}
\label{azim} \end{figure}
Figure \ref{azim} shows the azimuthal angle dependence of the RXS spectra.
As clear from Eqs.~(\ref{f100}), (\ref{f001}), and (\ref{f011}),
the peaks belonging to the same scattering vectors show the same
dependence.  In the (100) and (001) reflections, 
the intensities are proportional to $\cos^2\psi$ for the $\sigma\to\pi'$ 
channel, with the condition $E^{\mbox{\scriptsize in}}\parallel b$ at $\psi=0$.
In the the (011) reflection, the intensities are proportional to
$\sin^2 2\psi$ for the $\sigma\to\sigma'$ channel, with the condition
$E^{\mbox{\scriptsize in}}\parallel a$ at $\psi=0$, 
while the dependence is rather complicated for the $\sigma\to\pi'$ channel.  
The calculated results are in agreement with the those of YVO$_3$\cite{murakami00b}
(and also YTiO$_3$\cite{Nakao00}).

\begin{figure}
\begin{center}
\leavevmode
\epsfxsize=7cm \epsfbox{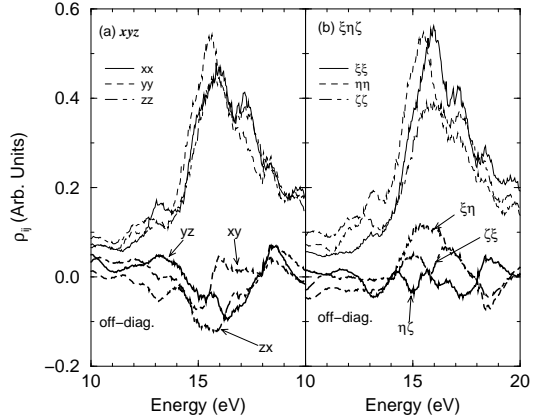}
\end{center}
\caption{
The $4p$ density matrix $\rho^{(1)}_{mn}(\varepsilon)$ 
at the Ti 1 site, as a function of energy $\varepsilon$,
(a) in the $xyz$ coordinate system, and (b) in the $\xi\eta\zeta$ coordinate
system. The thick lines represent the off-diagonal elements,
and the thin lines represent the diagonal elements.
The origin of energy is at the Fermi level.
}
\label{dmatrix} \end{figure}
To see more closely how the $4p$ states are modified,
we examine the $4p$ density matrix $\rho_{mn}^{(1)}(\varepsilon)$
at the Ti 1 site, which is shown in Fig. \ref{dmatrix}.
The left panel shows the representation in the $xyz$ system.
Although the diagonal elements are larger than the off-diagonal elements,
the former quantity does not contribute to the RXS intensity. 
The right panel shows the representation in the local $\xi\eta\zeta$
coordinate system. The off-diagonal elements remain comparable
to those in the $xyz$ system, indicating that the $\xi\eta\zeta$ axes
are still deviating from the principal axes of the density matrix.
Since the off-diagonal elements are fluctuating with changing energy,
the principal axes may also be fluctuating.
One may think that the principal axes of the density matrix are 
related to the direction of the $3d$ orbital polarization.
However, this is not the case in the present calculation,
since the $3d$ orbital polarization has little effect on the $4p$ states 
within the muffin-tin approximation.
We interpret this as the sensitiveness of the $4p$ states to the tilt of
the TiO$_6$ octahedra which makes the local $\xi\eta\zeta$ coordinate axes
no longer principal axes.

\section{Concluding remarks}
We have studied the effect of the crystal structure to the RXS spectra
of YTiO$_3$, using the band structure calculation
within the LDA. The calculated spectra are mainly compared with
the experimental data of YVO$_3$, since the unpublished data of YTiO$_3$ 
are similar to those of YVO$_3$. This similarity is consistent with 
the mechanism of the distorted crystal structure.
The obtained spectra as a function of photon energy are 
in agreement with the experiments of YVO$_3$ (and also YTiO$_3$).
Since the muffin-tin approximation averages the Coulomb interaction 
between the $4p$ states and the $3d$ states, the anisotropy of the
Coulomb interaction is eliminated in the calculation of the RXS spectra.
We have clearly shown that along with the JTD the tilts of
the transition-metal oxide octahedra need to be taken into account.
The agreement between the calculation and the experiment suggests 
that the RXS intensity originates mainly from the distorted crystal 
structure, casting doubt on a prevailing assertion that the RXS is a direct
observation of the orbital order.
We hope that the present work could accelerate further study 
on the relation between the RXS and the orbital order.

In spite of the above findings, there are several points to be made clear. 
First, since the crystal structure includes both the JTD and the tilt of the
TiO$_6$ octahedra, it is not clear which effect is important.
Closely related to this question, the crystal structure of YVO$_3$ changes
into a state of a very small JTD for $T > 100$ K (the tilt is not changed
so much). According to this change,
the intensities of peaks a and c were found to become very small,
while those of other peaks were found to remain similar.
This suggests that the effect of the JTD on the RXS spectra is different 
from that of the tilt of the octahedra. For clarifying this point
and for a direct comparison with the experiment, the calculation of
the RXS spectra in the two cases of the atomic coordinates for YVO$_3$ 
is now under progress.
Second, the intensities in the (100) reflection are $1/5\sim 1/10$ smaller 
than those in the other reflections in the experiment of YVO$_3$ 
(and YTiO$_3$). Note that the samples are different for different reflections
in the experiment. This may make the comparison of the data between different
reflections less accurate.\cite{com3}
In any event, this behavior is not reproduced by the present calculation.
It might improve the situation to go beyond the muffin-tin approximation 
in which the potential is crudely treated as a constant
in the interstitial region.
Third, in addition to the RXS spectra mentioned above, 
another small peak has been observed about 10 eV below the main $K$ edge 
peak. Since its azimuthal angle dependence is nearly the same as 
in other peaks,\cite{Nakao00} this pre-edge peak must come from 
the dipolar process. 
The presence of such pre-edge peaks has been predicted 
for LaMnO$_3$.\cite{Taka2}
This study suggests that the pre-edge peak arises from the $p$ symmetric 
states (centered at a Ti site) hybridizing to neighboring $3d$ states
in the intermediate states.
Since the orbital polarization is obtained very small in the present 
calculation, we are unable to handle the pre $K$ edge structure.
The LDA$+U$ method may be suitable to treat this problem.

\acknowledgments

We would like to thank H. Nakao and Y. Murakami for valuable
discussion and showing their data prior to publication.
MT wishes to thank H. Akai for providing the KKR code.
This work was partially supported by a Grant-in-Aid for Scientific
Research from the Ministry of Education, Culture, Sports, Science, 
and Technology, Japan.

\def\vol(#1,#2,#3){{\bf #1} (#2) #3}


\begin{references}

\bibitem{Murakami98a} Y. Murakami, H. Kawata, M. Tanaka, T. Arima, Y. Moritomo, and Y. Tokura:
                      Phys. Rev. Lett. \vol(80,1998,1932). 

\bibitem{Murakami98b} Y. Murakami, J. P. Hill, D. Gibbs, M. Blume, I. Koyama,
                      M. Tanaka, H. Kawata, T. Arima, Y. Tokura,
                      K. Hirota and Y. Endoh: 
                      Phys. Rev. Lett. \vol(81,1998,582). 

\bibitem{murakami99}  Y. Endoh, K. Hirota, S. Ishihara, S. Okamoto, Y. Murakami, A. Nishizawa,
                      T. Fukuda, H. Kimura, H. Nojiri, K. Kaneko, and S. Maekawa:
                      Phys. Rev. Lett. \vol(82,1999,4328).

\bibitem{murakami99b} K. Nakamura, T. Arima, A. Nakazawa, Y. Wakabayashi, and Y. Murakami:
                      Phys. Rev. B  \vol(60,1999,2425).   

\bibitem{murakami99c} M. von Zimmermann, J.P. Hill, D. Gibbs, M. Blume, D. Casa, B. Keimer, Y. Murakami, Y. Tomioka, and Y. Tokura:
                      Phys. Rev. Lett. \vol(83,1999,4872) 

\bibitem{murakami00b} M. Noguchi, A. Nakazawa, T. Arima, Y. Wakabayashi, H. Nakao, and Y. Murakami:
                      Phys. Rev. B \vol(62,2000,R9271)    

\bibitem{Nakao00}     H. Nakao, Y. Wakabayashi, T. Kiyama and Y. Murakami:
                      unpublished.                        

\bibitem{vettier99}   L. Paolasini, C. Vettier, F. de Dergevin, F. Yakhou, D. Mannix, A. Stunault, W. Neubeck, M. Altarelli, M. Fabrizio,
                      P. A. Metcalf, and J. M. Honig:
                      Phys. Rev. Lett. \vol(82,1999,4719).

\bibitem{murakami00}  
Y. Tanaka, T. Inami, T. Nakamura, H. Yamauchi, H. Onodera, K. Ohoyama,
and Y.Yamaguchi: J. Phys. Condens. Matter \vol(11,1999,L505);
K. Hirota, N. Oumi, T. Matsummura, H. Nakao, Y. Wakabayashi, Y. Murakami, and Y. Endoh:
                      Phys. Rev. Lett. \vol(84,2000,2706). 

\bibitem{Kanamori}    J. Kanamori:
                      J. Appl. Phys. Suppl. \vol(31,1960,14S).

\bibitem{Ishihara}   S. Ishihara and S. Maekawa:
                      Phys. Rev. Lett. \vol(80,1998,3799);
                      Phys. Rev. \vol(B 58,1998,13449).

\bibitem{Elfimov}     I. S. Elfimov, V. I. Anisimov and G. Sawatzky:
                      Phys. Rev. Lett. \vol(82,1999,4264).

\bibitem{Benfatto}    M. Benfatto, Y. Joly and C. R. Natoli: 
                      Phys. Rev. Lett. \vol(83,1999,636). 

\bibitem{Taka1}       M. Takahashi, J. Igarashi and P. Fulde:
                      J. Phys. Soc. Jpn. \vol(68,1999,2530).

\bibitem{Sawada98}    H. Sawada and K. Terakura:
                      Phys. Rev. B \vol(58,1998,6831).

\bibitem{Daivid79}    D. A. MacLean, Hok-Hamng, and J. E. Greedan:
                      J. Solid State Chem. \vol(30,1979,35)   

\bibitem{com1}        Generally, the TiO$_6$ octahedra are rotated. We simply call this rotation as ``tilt".

\bibitem{Kawano94}    H. Kawano, H. Yoshizawa and Y. Ueda:     
                      J. Phys. Soc. Jpn. \vol(63,1994,2857).

\bibitem{Benedetti00} P. Benedetti, J. van den Brink, E. Pavarini, A. Vigliante, and P. Wochner:
                      cond-mat/0012084

\bibitem{com2}
The JTD effect is roughly proportional to the square of the magnitude of 
the distortion.

\bibitem{Iga1}        J. Igarashi and K. Hirai:
                      Phys. Rev. B \vol(50,1994,17820), \vol(53,1996,6442).

\bibitem{Iga2}        J. Igarashi and M. Takahashi: J. Phys. Soc. Jpn.
                      \vol(69,2000,4087). 

\bibitem{Iga3}        J. Igarashi and M. Takahashi: Phys. Rev. B, in press.

\bibitem{Cowan}       R. Cowan, {\em The Theory of Atomic Structure and Spectra}
                      (University of California Press, Berkeley, 1981).

\bibitem{Goral82}     J. P. Goral, J. E. Greedan and D. A. MacLean:
                      J. Solid State Chem. \vol(43,1982,244).  

\bibitem{asano95}     H. Fujitani and S. Asano:
                      Phys. Rev. B \vol(51,1995,2098).

\bibitem{zeller}      R. Zeller : J. Phys. C: Solid State Phys. \vol(20,1997,2347).

\bibitem{com3}
The raw data are devided by the intensity of the (200) fundamental reflection
in each sample in the comparison.

\bibitem{Taka2}       M. Takahashi, J. Igarashi and P. Fulde:
                      J. Phys. Soc. Jpn. \vol(69,2000,1614).


\end{references}
\end{document}